\documentclass[namedreferences,hyperref,optionalrh]{spr-sola}
\usepackage{graphicx}        % For eps figures, newer & more powerfull
\usepackage{color}           % For color text: \color command
\renewcommand{\vec}[1]{{\mathbfit #1}}

\chardef\us=`\_

\begin{document}

%%\begin{article}
\begin{opening}

\title{Data-Constrained Magnetohydrodynamics Simulation of a Confined 
X-class Flare in NOAA Active Region 11166}

\author[addressref={aff1},corref,email={sainisanjay35@gmail.com}]{\inits{S.}\fnm{Sanjay}~\lnm{Kumar}\orcid{0000-0003-4578-6572}}
\author[addressref={aff1}]{\inits{P.}\fnm{Pawan}~\lnm{Kumar}}
\author[addressref={aff1}]{\inits{S.}\fnm{Sadashiv}~\lnm{}}
\author[addressref={aff2},corref]{\inits{S.}\fnm{Sushree S.}~\lnm{Nayak}\orcid{0000-0002-4241-627X}}
\author[addressref={aff3}]{\inits{S.}\fnm{Satyam}~\lnm{Agarwal}\orcid{0000-0002-0833-9485}}
\author[addressref={aff4,aff5}]{\inits{A.}\fnm{Avijeet}~\lnm{Prasad}\orcid{0000-0003-0819-464X}}
\author[addressref={aff3}]{\inits{R.}\fnm{Ramit}~\lnm{Bhattacharyya}\orcid{0000-0003-4522-5070}}
\author[addressref={aff6}]{\inits{R.}\fnm{Ramesh}~\lnm{Chandra}\orcid{0000-0002-3518-5856}}
%%%\author[addressref={aff6}]{\inits{U.}\fnm{Upendra}~\lnm{Kushwaha}\orcid{0000-0003-0819-464X}}

\address[id=aff1]{Department of Physics, Patna University, Patna 800005, India}
\address[id=aff2]{Center for Space Plasma and Aeronomic Research, the University of Alabama in Huntsville, AL, 35899, USA}
\address[id=aff3]{Udaipur Solar Observatory, Physical Research Laboratory,
Dewali, Bari Road, Udaipur 313001, India}
\address[id=aff4]{Institute of Theoretical Astrophysics, University of Oslo, PO Box 1029 Blindern, 0315 Oslo, Norway}
\address[id=aff5]{Rosseland Centre for Solar Physics, University of Oslo, PO Box 1029 Blindern, 0315 Oslo, Norway}
\address[id=aff6]{Department of Physics, DSB Campus, Kumaun University, 263001 Nainital, India}
%%\address[id=aff4]{Department of Physics, University of Allahabad, Prayagraj (Allahabad)- 211002}

\runningauthor{Kumar et al.}
\runningtitle{MHD Simulation of a confined flare }

\begin{abstract}
{In this paper, we present a magnetohydrodynamics simulation of 
NOAA active region 11166 to understand the origin of a confined 
X-class flare that peaked at 23:23 UT on 2011 March 9. The simulation is 
initiated with a magnetic field extrapolated from the corresponding 
photospheric magnetogram using a non-force-free-field extrapolation technique. 
Importantly, the initial magnetic configuration identifies three-dimensional (3D) magnetic nulls and quasi-separatrix layers (QSLs), which nearly agree with the bright structures appeared in multi-wavelength observations.
The Lorentz force associated with the extrapolated field self-consistently generates the dynamics that leads to the magnetic reconnections at the 3D nulls and the QSLs. These reconnections are found to contribute to the pre-flare activities and, ultimately, lead to the development of the flare ribbons. Notably, the anchored spine of the 3D null and the complete absence of flux rope in the flaring region are congruent with the conﬁned nature of the ﬂare. Furthermore, the simulation also suggests the role of reconnections at the 3D null with an open spine in the onset of a jet away from the flaring site. }
\end{abstract}


\end{opening}
%-------------------------------------------------

\section{Introduction}
\label{S-Introduction} 
Solar flares are the manifestation of an explosive and sudden release of excess magnetic energy from magnetically dominated solar coronal plasma \citep{Shibata2011}. The underlying mechanism for their occurrence is considered to be magnetic reconnection that rapidly generates plasma flow, accelerated charged particles, and radiation from the magnetic energy \citep{2023AdSpR..71.1856P}. However, the physics of magnetic reconnection is still not fully understood, particularly in the complex three-dimensional (3D) coronal magnetic configuration \citep{Shibata2011, 2023AdSpR..71.1856P}.

Notably, it has been shown that sometimes, the flares can be related to coronal mass ejections (CMEs) --- leading to their nomenclature ``eruptive flares" {\citep{2001ApJ...559..452Z, 2008ApJ...673L..95T}}. In such cases, it is considered that flares and CMEs may be different manifestations of the underlying magnetic reconnection process in the corona \citep{2002A&ARv..10..313P, 2006ApJ...649.1100Z, 2008ApJ...673L..95T}. Phenomenologically, the typical eruptive flares are explained by a standard flare model a.k.a CSHKP model \citep{1964NASSP..50..451C, 1966Natur.211..695S, 1974SoPh...34..323H, 1976SoPh...50...85K}. The model mainly considers the magnetic configuration in the form of a simple bipolar region in which a magnetic flux rope (representing a filament) is proposed to be rising ---
stretching the overlying field lines and producing a current sheet below the flux rope. Consequently, parallel ribbons of the flares are formed due to reconnection at the current sheet and the erupting rope leads to the associated CME. 

Various studies with multi-wavelength observations and magnetic topology have shown different complex flaring processes producing flare ribbons with a variety of geometries that do not match the ones proposed in the standard model {\citep{1996SoPh..168..115M, 2000ApJ...540.1126A, Masson2009, 2012ApJ...760..101W, 2020SoPh..295...75D}}. In particular, the onset of the flares without any eruptions, also known as confined flares, is considered to be difficult to understand from the CSHKP model \citep{1997A&A...325.1213S, 2004A&A...423.1119B, 2012ApJ...749..135J, 2015A&A...574A..37D} and requires further studies that involve the evolution of coronal magnetic configuration leading to reconnection.
Notably, previous studies by \citet{Masson2009, 2016ApJ...828...62J, prasad+2018apj} based on magnetohydrodynamics (MHD) simulation attempted to explain the occurrence of such types of confined flares, lacking a flux rope in the flaring region. In particular, \citet{Masson2009, prasad+2018apj} detected 3D magnetic null points in the flaring region which lead to the null-point reconnections and formation of the circular ribbons. \citet{2016ApJ...828...62J} suggested that the formation of large-scale current-sheet in a sheared arcade and consequent magnetic reconnection are responsible for the parallel ribbons in a confined flare.   

%%% insert studies related to confined flares %%%%%%%%

In this paper, we select a confined flaring event having a complex system of flare ribbons that occurred in NOAA active region (AR) 11166 on 2011 March 9. Relevantly, previous studies of this AR \citep{2012ApJ...761...86V, 2014ApJ...792...40V, li2019} have been carried-out using observations and numerical modeling with the nonlinear-force-free-field (NLFFF) {\citep{2008JGRA..113.3S02W}}. These studies focused primarily on the helicity ejection and the role of magnetic topology in the ribbon emission. However, to explore the flare dynamics, we employ the data-constrained MHD simulation initiated from an extrapolated magnetic field. To construct the extrapolated magnetic configuration of the AR, here we have considered a non-force-free-field (NFFF) extrapolation model {\citep{hu&dasgupta2008soph, 2008ApJ...679..848H, 2010JASTP..72..219H}}. Crucially, the extrapolated magnetic configuration supports a non-zero Lorentz force. The force is important to trigger the flare dynamics in the simulation. Relevantly, the Lorentz force is considered to be a key parameter to understanding the triggering mechanisms of the coronal transients \citep{2019ApJ...885L..17S}. Recent data-constrained MHD simulations driven by NFFF extrapolations are proven to be successful in reproducing active region dynamics that leads to flaring events as well as other phenomena such as coronal jets, coronal dimmings, and flux rope formation {\citep{prasad+2018apj, 2019ApJ...875...10N, prasad_2020, 2021PhPl...28b4502N, 2023A&A...677A..43P}}. Relevantly, the NLFFF extrapolation initiated MHD simulations are also found to be successful in simulating the dynamics of the similar transient events in the solar corona \citep{2013ApJ...771L..30J, 2015ApJ...803...73I, jiang+2016nat, 2016ApJ...817...43S, 2023ApJ...944..179F, 2023ApJ...949....2Z}.

The NFFF extrapolation of AR 11166 suggests the presence of several magnetic features, such as 3D magnetic nulls and quasi-separatrix layers (QSLs). The null is the location where magnetic field $\textbf{B}=0$, whereas the QSL is the site where a sharp gradient in magnetic connectivity is found {\citep{1999A&A...351..707T,Titov2002, aulanier2006SoPh}}.
They are considered to be preferable sites for magnetic reconnection {\citep{aulanier2006SoPh, 2021SoPh..296...26K}}. Notably, the morphology of these features in the extrapolated field is similar to the co-temporal brightenings identified in multi-wavelength observations of the event. The simulated dynamics documents the onset of magnetic reconnection at the 3D null with an anchored spine and the slipping magnetic reconnections {\citep{aulanier2006SoPh}} at the QSLs in the flaring region --- causing the pre-flare activities and the different-shaped ribbon emission. Interesting is the slipping reconnections at one of the QSLs, explaining the brightening observed during the initiation of the flare. The simulation also identifies the reconnections at the other 3D null with an open spine to be central in a jet situated at some distance from the flaring site.  Notably, we find that the anchoring of the spine corresponding to the 3D null in the flaring region can be responsible for the confined nature of the flare. The absence of the flux rope in the region further corroborates the finding. 

In the following, Section~\ref{obs-extra} briefly discusses the important observations of the flaring event and provides the key magnetic features of the initial non-force-free-field extrapolated configuration. Section~\ref{mhd-sim} presents the governing MHD equations and the numerical model, along with the computational setup used for the MHD simulation. In Section~\ref{results}, we discuss the simulation results and their relations to the multi-wavelength observations. The findings of the paper are highlighted in Section~\ref{summary}.

\section{Observations and Extrapolation} %%%%%%%%%%%%%%%%%%%%%%%%%%%%%%%%%%%%%%%%
\label{obs-extra}      

\subsection{Observational features of the confined flare}
\label{features}
\begin{figure}[ht!]
\centerline{\includegraphics[width=\textwidth]{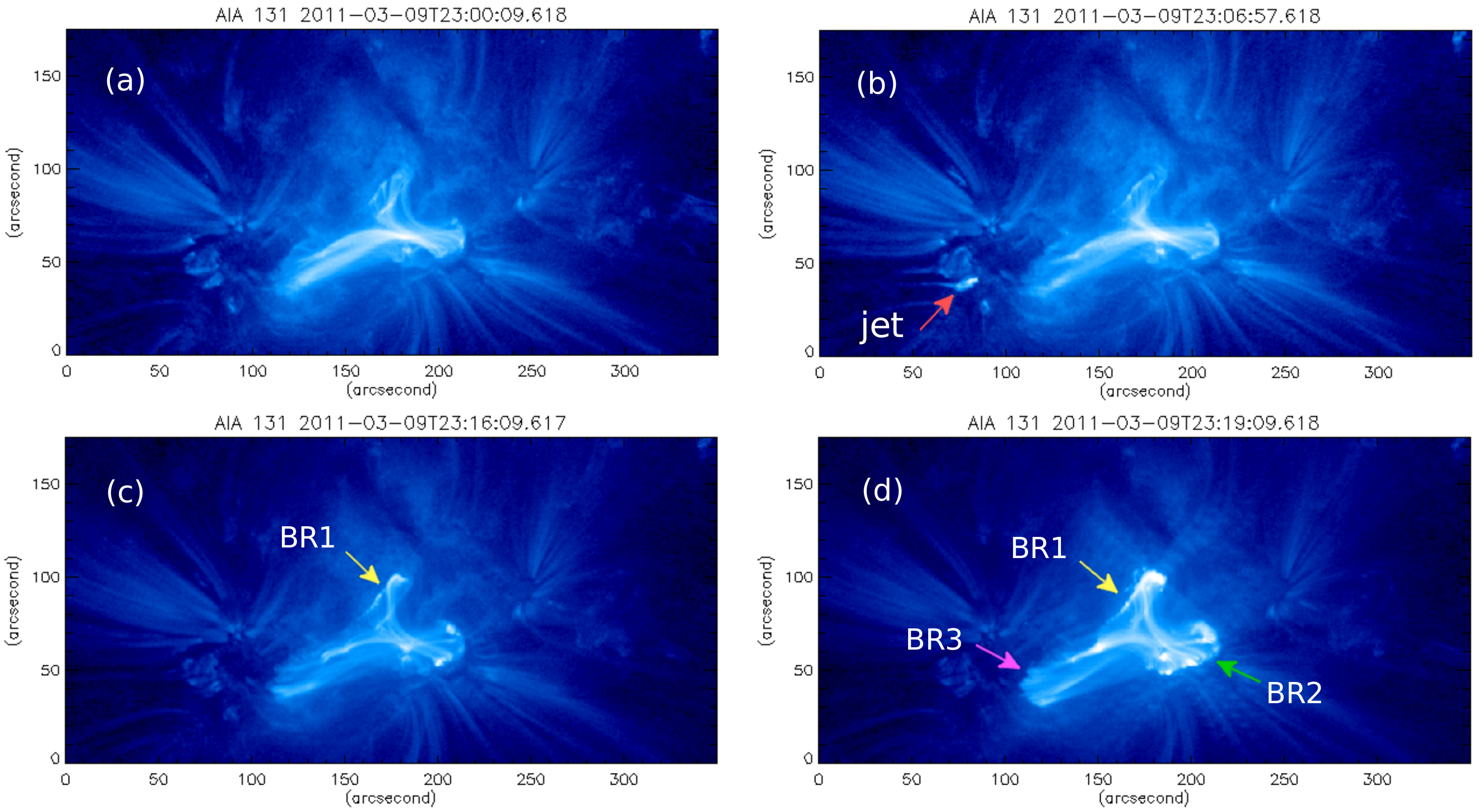}}
\caption{The X1.5 flare in 131 {\AA} from t $\approx$ 23:00 UT to $\approx $ 23:19 UT. Panel (b) shows the jet brightening. Panel (c) highlights the appearance of a flare brightening, which enhances and moves in a downward direction (see yellow arrows). The green and pink arrows mark a circular and relatively faint diffused brightening, respectively.   An animation of this figure is available. }
\label{obs-131}
\end{figure}

\begin{figure}[ht!]
\centerline{\includegraphics[width=\textwidth]{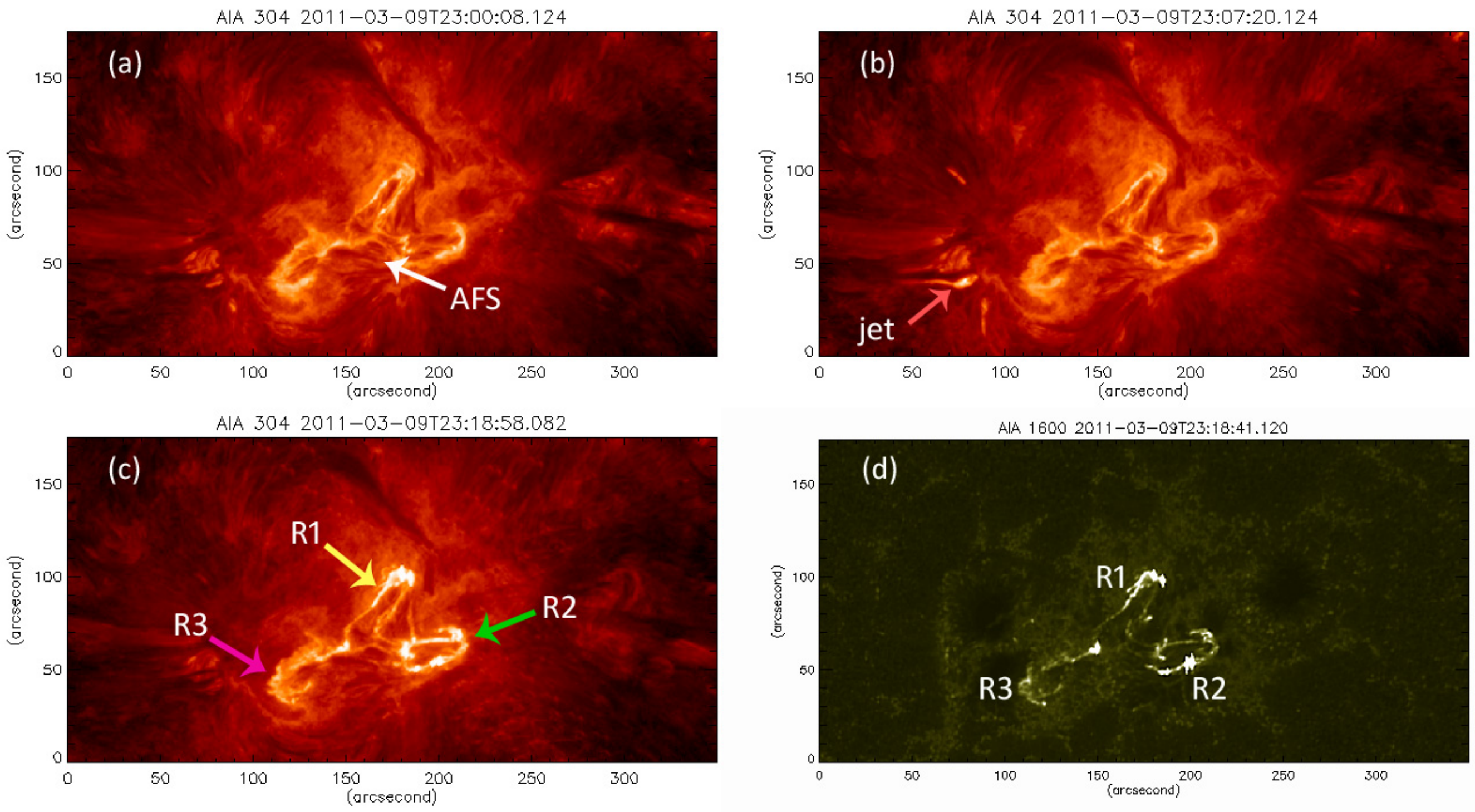}}
\caption{Panels (a)-(c) show the flare evolution in 304 {\AA} along with the jet (panel (b)). The presence of AFS is indicated by a white arrow in panel (a).  
Panels (c) and (d) illustrate the flare ribbons (marked by R1, R2, and R3) in 304 {\AA} and 1600 {\AA}, respectively. 
 An animation of panels (a)-(c) is available.}
\label{obs-304}
\end{figure}

The confined X1.5 class flare is produced near the center of the solar disk in NOAA AR 11166 on 2011 March 9. The multi-wavelength study of the flare is already being conducted by \citet{li2019}. The flare is found to be initiated at 23:13 UT and peaked at 23:23 UT in GOES observations. Figures \ref{obs-131} and \ref{obs-304} illustrate some of its key observational features obtained from the Atmospheric Imaging Assembly \citep[AIA,][]{Lemen2012} onboard Solar Dynamics Observatory \citep[SDO,][]{Pesnell2012} which observes the Sun in EUV and UV wavelengths with a pixel size of 0.6$''$ and with a temporal resolution of 12s.  

Figure \ref{obs-131} illustrates the pre-flare evolution in AIA 131 {\AA} starting from 23:00 UT (panel (a)). Interestingly, around 23:07 UT, a localized brightening (red arrow in panel (b)) is observed having structures similar to the base and spire of a typical jet \citep{2010ApJ...720..757M}, which indicates the occurrence of a jetting phenomenon. Further in time, the elongated flare brightening (BR1) starts to appear (yellow arrow in panel (c)), and also observable is its movement in the downward direction (yellow arrow in panel (d)). Importantly, we also observed the circular flare brightening (BR2), marked by a green arrow in panel (d). In addition, a relatively faint brightening (BR3) is observed having a semi-circular kind of shape (pink arrow in panel (d)). In Figure \ref{obs-304}, we present observations of the flare in AIA 304 {\AA} (panels (a) to (c)) and 1600 {\AA} (panel (d)) channels. Panel (b) marks the presence of a distant localized brightening (red arrow) similar to the one observed in 131 {\AA} (Figure \ref{obs-131}(b)). 
In 304 {\AA} observations, we can identify the elongated flare ribbon (R1) that extends in the downward direction (yellow arrow in panel (c)) and the circular-shaped ribbon (R2), depicted by a green arrow in panel (c). Additionally, the semi-circular-shaped flare ribbon (R3) having relatively less brightness is found to form (pink arrow in panel (c)). Overall, Figure \ref{obs-304}(d) also identifies three different types of flare ribbons in 1600 {\AA}: an elongated ribbon, a circular ribbon, and a less-bright semi-circular ribbon. The three ribbons are marked by R1, R2, and R3 respectively. Moreover, noticeable is the possible presence of an arch filament system \citep[AFS,][]{2002A&A...391..317M} in AIA 304 {\AA} channel (white arrow in Figure \ref{obs-131}(a)). The AFS is found to be stable and its presence did not affect the flare evolution \citep{li2019}.

%%%%added by sushree%%%%%%

\subsection{The Non-Force-Free-Field Extrapolation of AR11166}
\label{nfff}
To investigate the onset of the confined flare, we first construct the initial magnetic field topology of AR11166 using 
the non-force-free-field extrapolation model \citep{2008ApJ...679..848H,2010JASTP..72..219H}. The model is based on the minimum dissipation rate (MDR) principle, which ensures the relaxed state of the plasma to have the minimum dissipation rate \citep{2004PhPl...11.5615B, 2007SoPh..240...63B}.
As a result, the magnetic field $\textbf{B}$ is the solution of the double-curl Beltrami equation and can be written as \citep{2007SoPh..240...63B}
\begin{equation}
\mathbf{B} = \mathbf{B_1}+\mathbf{B_2}+\mathbf{B_3}; \quad \nabla\times\mathbf{B_i}=\alpha_i\mathbf{B_i}
\label{e:b123}
\end{equation}
with $i=1,2,3$. Here, each ${\bf{B}}_i$ represents a linear-force-free field (LFFF) with corresponding twist parameter $\alpha_i$. In the practical NFFF extrapolation technique, without loss of generality, we choose $\alpha_1\ne\alpha_3$ and $\alpha_2 = 0$ that makes $\mathbf{B_2}$ a potential field. Then each ${\bf{B}}_i$ with given $\alpha_i$ can be obtained into the computational volume by using a fast Fourier transform method.
Subsequently, an optimal pair $\alpha=\{\alpha_1, \alpha_3\}$ is attained by an iterative scheme which is based on the minimization of the average deviation between the observed ($\mathbf{B}_t$) and the calculated ($\mathbf{b}_t$) transverse field on the photospheric boundary. The deviation is calculated by the following metric \citep{2010JASTP..72..219H,prasad+2018apj}:
\begin{equation}
E_n =\left(\sum_{i=1}^M |\mathbf{B}_{t,i}-\mathbf{b}_{t,i}|\times|\mathbf{B}_{t,i}|\right)/\left(\sum_{i=1}^M |\mathbf{B}_{t,i}|^2\right)
\label{en}
\end{equation}
\noindent where $M$ represents the total number of grid points on the transverse plane. For further details on the NFFF extrapolation model, the readers are referred to \citet{hu&dasgupta2008soph,2008ApJ...679..848H,2010JASTP..72..219H}.

The extrapolated field supports a non-zero Lorentz force, crucial for driving the plasma in the presented MHD simulation. 
 Moreover, the use of the NFFF model is also in line with the 
 study by \citet{2001SoPh..203...71G}, that examined the variation of plasma-$\beta = \frac{p}{p_{mag}}$ (where $p$ and $p_{mag}$ represent the plasma and magnetic pressure respectively) over the active regions. The study suggests that the force-free approximations may not hold well near the photosphere, especially away from the sunspots. \citet{1995ApJ...439..474M} have also shown the absence of the force-free equilibrium lower than 400 km in the solar atmosphere. Relevantly, in standard force-free extrapolation models, a ``preprocessing" technique is often utilized to remove the force in the vector magnetogram and achieve 
a suitable boundary condition for the extrapolations {\citep{2008JGRA..113.3S02W, 2012LRSP....9....5W}}.
 In comparison, the NFFF extrapolation technique does not require such apriori preprocessing. It is noteworthy that the recent studies using MHD simulation initiated with the NFFF successfully explained various transient events in the active regions such as flares, coronal jets, and coronal dimmings {\citep{prasad+2018apj, 2019ApJ...875...10N, prasad_2020, 2023A&A...677A..43P}} and established the model as an alternative to the routinely utilized nonlinear-force-free models {\citep{2022SoPh..297...91A}}.

\begin{figure}[ht!]
\centerline{\includegraphics[width=\textwidth]{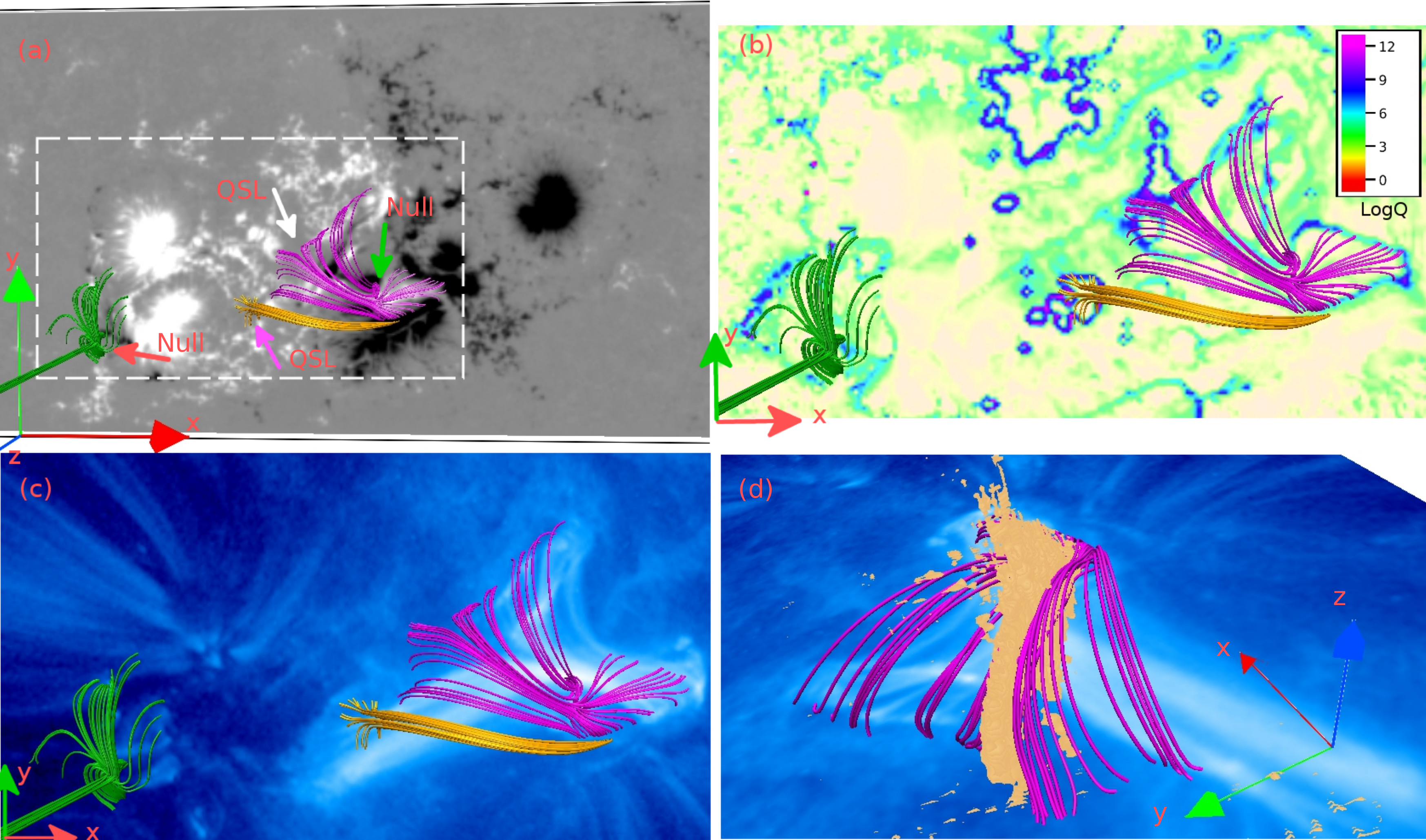}}
\caption{Panel (a) shows the extrapolated field lines, with Bz on the bottom boundary in the grayscale. In panel (a), the presence of 3D nulls is marked by red and green arrows. QSLs are indicated by white and pink arrows. Panel (b) plots the LogQ at the bottom boundary for the boxed part (white rectangle in panel (a)) overlaid with the field lines. Panel (c) depicts the field lines for the boxed part overplotted with 131 {\AA} image. Isosurface of LogQ with isovalue 9 (in color buff orange) is plotted with pink-colored magnetic loops (panel (d)).}
\label{extrapol}
\end{figure}

%%%%%%%%%%%%%%%%%%%%%%%%%%%%%%%
For performing the NFFF extrapolation, Helioseismic and Magnetic 
Imager \citep[HMI,][]{2012SoPh..275..229S} magnetogram is taken from 
the `hmi.sharp\_cea\_720s' data series that provides photospheric vector magnetograms of the Sun with a temporal cadence of 12 minutes and a pixel resolution of $0.5''$. To obtain the magnetic field on a Cartesian grid, the magnetogram is initially remapped onto a Lambert cylindrical equal-area (CEA) projection and then transformed into heliographic coordinates \citep{1990SoPh..126...21G}.
We selected the magnetogram at 23:00 UT for the extrapolation. The magnetogram cut-out was originally 700 $\times$ 348 pixels along $x$ and $y$ directions in a Cartesian coordinate system. We have performed the NFFF extrapolation for a re-scaled size of 350 $\times$ 174 grid points along $x$ and $y$ to reduce the computational cost. The physical extent of the computational box is $\approx$ 252 Mm in $x$ and $\approx$ 126 Mm in both $y$ and $z$ directions.

The extrapolated magnetic field lines are plotted in Figure \ref{extrapol}. In the extrapolated magnetic field, we identify two 3D magnetic nulls, as depicted in panel (a) of Figure \ref{extrapol}. The nulls are identified using the procedure documented by \citet{2020ApJ...892...44N}. Succinctly, the procedure is based on the construction of a Gaussian indicator,
\begin{equation}
\label{nullfunc}
\chi (x,y,z)= exp\left[-\sum_{i=x,y,z}\frac{(B_i(x,y,z)-B_0)^2}{{d_0}^2}\right].
\end{equation}
\noindent where $d_0$ and $B_0$ represent the Gaussian width and a particular isovalue of $B_i$, respectively. 
With a choice of $B_0 \approx0$, the function $\chi(x,y,z)$
takes significant values only if $B_i\approx B_0$ for each $i$. 
The 3D nulls are then the points where the three isosurfaces with isovalues $B_i=B_0$ intersect. The procedure is used successfully to locate nulls in the analytical fields \citep{2020ApJ...892...44N, 2021SoPh..296...26K} and in the extrapolated fields \citep{2019ApJ...875...10N, 2022ApJ...925..197B}. The null (marked by a green arrow) located in the flaring region is characterized by the anchored spine and the dome-shaped fan and is represented by the pink field lines. 
The null is located at a height $\approx$ 1.04 Mm. The other null (highlighted by a red arrow) having an open spine (in the computational domain) and a dome-shaped fan is shown by green field lines and is situated away from the flaring site. 
The height of this null is $\approx$ 1.43 Mm. The null is approximately co-spatial with the jet location (see Figures \ref{obs-131}(b) and \ref{obs-304}(b)). 
Noteworthy is a spreading of the pink field lines (located in the vicinity of the spine of the null) from each other in the flaring region --- suggesting the presence of a QSL (marked by a white arrow in panel (a)). Moreover, the yellow field lines also fan out in this region, which indicates the existence of another QSL (marked by a pink arrow in panel (a)).

\begin{figure}[ht!]
\centerline{\includegraphics[width=\textwidth]{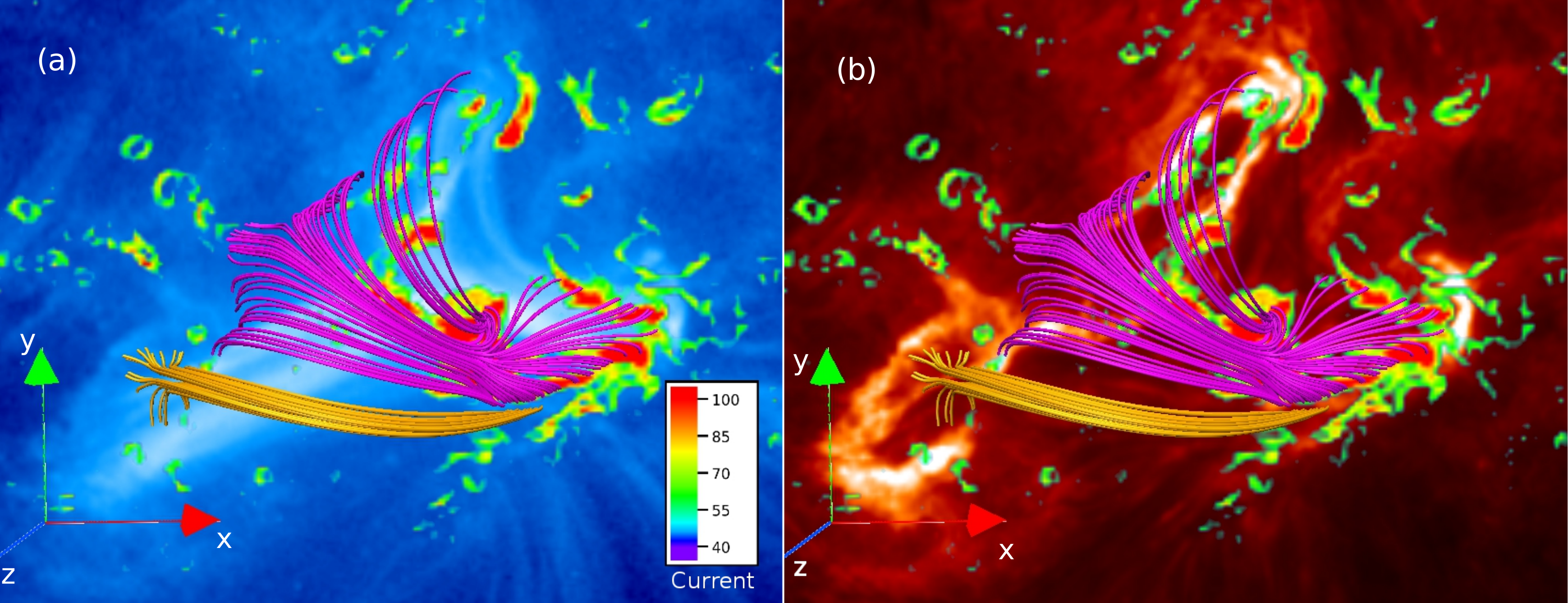}}
\caption{The figure shows the extrapolated field lines in the flaring region overlaid with the images in 131 {\AA} (panel (a)) and 304 {\AA} (panel (b)). We also overplot the figure with the distribution of the current density on the background. }
\label{current}
\end{figure}

To further confirm the existence of the QSLs, Figure \ref{extrapol}(b) shows the field lines overlaid with the squashing factor \citep[Q,][]{liu-2016} at the bottom boundary. Notable is the higher Q values in the region where the spreading of the pink field lines takes place 
(white arrow in Figure \ref{extrapol}(a)) --- illustrating a QSL. Moreover, we also identify another region with high Q-values co-located with the fanning out yellow field lines (pink arrow in Figure \ref{extrapol}(a)), which demonstrates another QSL. Figure \ref{extrapol}(c) plots the extrapolated field lines with AIA 131 {\AA} image in the background at 23:00 UT. The AIA cutouts are also CEA projected and scaled according to the HMI pixel resolution. Importantly, the pink-colored field lines match morphologically well with the observed brightened loops before the flare. From a side viewpoint, it is evident that the spreading of the pink field lines and the brightening is co-spatial with the QSL (Figure \ref{extrapol}(d)). The presence of high Q values in the vicinity of the spreading of the field lines is evident in panel (d). To substantiate further, in Figure \ref{current}, we show the distribution of current density ($|\mathbf{J}|$) on the bottom boundary, overlaid with 131 {\AA} (panel (a)) and 304 {\AA} (panel (b)) images of the flaring region at 23:00 UT. Notable is the concentration of high values of currents in the bright regions --- pointing toward the role of the joule heating in these locations.

\begin{figure}[ht!]
\centerline{\includegraphics[width=\textwidth]{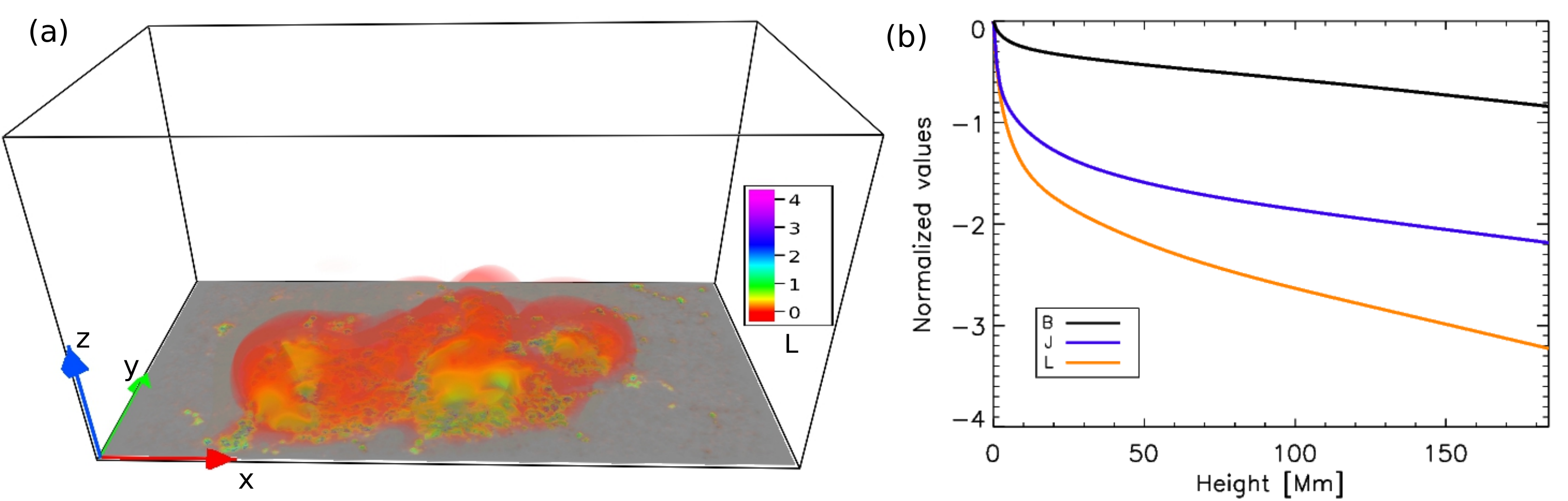}}
\caption{Panel (a) illustrates the direct volume rendering of the Lorentz force density in the units of $1.2 \times 10^{-5}$ dyne cm$^{-3}$.  Panel (b) depicts the logarithmic variation of the horizontally averaged magnetic field (B), the current density (J), and the Lorentz force density (L) with height. All of the plotted parameters are normalized to their maximum values.}
\label{lorentz}
\end{figure}

Figure \ref{lorentz}(a) depicts the distribution of the initial Lorentz force density in the computational domain. Notable is the presence of the force density mostly at the lower heights. The figure also documents the strong Lorentz force in the flaring region, which plays a key role in initiating the dynamical evolution. Figure \ref{lorentz}(b) shows the variation of horizontally averaged strength for the magnetic field, current density, and Lorentz force density with height. 
The figure demonstrates a faster decay of the Lorentz force density in comparison to the current density and the field strength. Also, noticeable is the fall of the Lorentz force by around two orders of magnitude at 10 Mm height.

\section{MHD simulation of AR11166} 
\label{mhd-sim}
\subsection{MHD equations and Numerical Model} 
Toward exploring the active region dynamics and associated topological changes in the magnetic field, we approximate the plasma to be an incompressible magnetofluid. Moreover, the magnetofluid is considered to be of uniform density, thermally homogeneous, and perfectly electrically conducting \citep{2017PhPl...24h2902K, 2021SoPh..296...26K}. Such a consideration suffices for investigating the initiation mechanism of the coronal transients, particularly the changes in field line connection due to reconnection, as exemplified in earlier works {\citep{2019ApJ...875...10N, 2022FrASS...939061K, 2023A&A...677A..43P}}. The MHD equations in the dimensionless form are given as: 

\begin{eqnarray}
\label{stokes}
& & \frac{\partial{\bf{v}}}{\partial t} 
+ \left({\bf{v}}\cdot\nabla \right) {\bf{ v}} =-\nabla p
+\left(\nabla\times{\bf{B}}\right) \times{\bf{B}}+\frac{\tau_\textrm{a}}{\tau_\nu}\nabla^2{\bf{v}},\\  
\label{incompress1}
& & \nabla\cdot{\bf{v}}=0, \\
\label{induction}
& & \frac{\partial{\bf{B}}}{\partial t}=\nabla\times({\bf{v}}\times{\bf{B}}), \\
\label{solenoid}
 & & \nabla\cdot{\bf{B}}=0, 
\label{e:mhd}
\end{eqnarray}

\noindent with strength of ${\bf{B}}$ and the plasma velocity ${\bf{v}}$ are being normalised by the average magnetic field strength ($B_0$) and  the Alfv\'{e}n speed 
($v_a \equiv B_0/\sqrt{4\pi\rho_0}$ with $\rho_0$ representing the constant mass density), respectively. The spatial scale, $L$, and the temporal scale, $t,$ are normalized by the length-scale of the vector magnetogram, ($L_0$), and the Alfv\'{e}nic transit time, ($\tau_a=L_0/v_a$).  While the plasma pressure $p$ is scaled with ${\rho {v_a}^2}$. 
Here, $\tau_\nu$ denotes the viscous diffusion time scale 
($\tau_\nu= L_0^2/\nu$), with $\nu$ being the kinematic viscosity. The pressure $p$ satisfies the equation

\begin{equation}
\nabla^2 \left(p+ \frac{v^2}{2}\right)=\nabla\cdot[(\nabla \times \vec{B})\times \vec{B}-(\nabla \times \vec{v})\times \vec{v}],
\label{en1}
\end{equation}

\noindent obtained by imposing the incompressibility (Equation \ref{incompress1}) on the momentum transport equation (Equation \ref{stokes}). Equation \ref{en1} is an elliptic partial differential equation and solved as a boundary value problem. It is important to note that, for incompressible flows, the pressure $p$ can not be related to density or temperature through an equation of state \citep{kajishima2017} --- making the system thermodynamically inactive to any pressure perturbation. In the absence of time derivative, it just adjusts itself instantaneously in a flow field to keep the velocity field divergence free --- amounting to an infinite sound speed \citep{2010PhPl...17k2901B}. Also notable is that, in the momentum transport equation, there is only the gradient of the pressure, but not the pressure itself. Hence, in incompressible flow solution, the actual value of the pressure is not important, only the changes of the pressure in space are important.

The solution of the MHD equations as the initial value problem is obtained from the well-established numerical model EULAG-MHD.
The model is well described in \citet{2013JCoPh.236..608S} and references therein. The advection scheme of the model is the spatio-temporally second-order accurate nonoscillatory forward-in-time multi-
dimensional positive deﬁnite advection transport algorithm,
MPDATA \citep{2006IJNMF..50.1123S}. An important feature of MPDATA and crucial in our simulation is its dissipative property, intermittent and adaptive, that
regularizes the under-resolved scales developed in the magnetic field by simulating reconnections --- imitating the operation of explicit subgrid scale turbulence models in the spirit of implicit large eddy simulation 
\citep[ILES,][]{grinstein2007book}. In our previous works \citep{prasad_2020, 2019ApJ...875...10N, 2022FrASS...939061K, 2023A&A...677A..43P}, we have extensively studied the dynamics of various active regions with the model and successfully explained the various events such as flares, coronal jets, and coronal dimming in the active regions.

\subsection{Computational Setup}
Here we conduct the MHD simulation in a domain with computational grid points  $350\times 174 \times 174$ for a physical domain spanning $\approx$ $[0,2]\times [0,1]\times [0,1]$ units in $x$, $y$, and $z$, respectively, where a unit length approximately corresponds to $126$ Mm. 
In the simulation, with the NFFF field (Figure \ref{extrapol}) as the initial magnetic field, the initial state is considered to be motionless ($\mathbf{v}=0$). With a minimal change in magnetic ﬂux at the boundary during the chosen confined flare, the normal components of the magnetic field ($\mathbf{B}$) and velocity field ($\mathbf{v}$) are fixed to their initial values at the boundaries. For the other components, we took all the variables by linearly extrapolating their values from the interior points in their spatial neighborhood \citep{prasad+2018apj, 2023A&A...677A..43P}. 
Further, to perform the simulation with optimal computational cost, we choose the non-dimensional constant $\tau_a / \tau_\nu$  to be $2 \times 10^{-4}$. The constant is around 15 times larger than its coronal value, which helps us to speed up the dynamical evolution without affecting the magnetic topology. 
With the Courant-Friedrichs-Lewy (CFL) stability condition \citep{1967IBMJ...11..215C}, the spatial unit step $\Delta x = 0.00576$ and time step (normalised by the Alfv\'{e}n  transit time $\tau_a \sim 30s$) $\Delta t = 5\times10^{-3}$. We carry out the simulation for 500 $\Delta t$ which is roughly 19 minutes of observation time. For a better comparison of the MHD simulation with the observations, in this paper, we present the time ($t$) in units of $2 \tau_a = 1$ minute in describing the simulation results.

With a choice of $\tau_a / \tau_\nu= 2 \times 10^{-4}$, the Alfv\'{e}n speed is $v_a \approx 6.6 \times 10^4$m/s. Under the incompressibility condition, in our simulation, the pressure is not determined thermodynamically using an equation of state and changes to ensure the divergence free-constraint (Equation \ref{incompress1}) --- an assumption also used in other works \citep{dahlburg+1991apj,aulanier+2005aa}.   
Although the compressibility is crucial to explore the thermodynamics of the coronal loops  \citep{ruderman&roberts2002apj}, here we only focus on examining the changes in magnetic topology during the confined flare and the jet in a thermodynamically inactive ambient state.

\section{Results of MHD Simulation}
\label{results}
In the simulation, the initial Lorentz force inherent to the NFFF field pushes the plasma from the motionless state and triggers the dynamical evolution of the active region. From panel (b) of Figures \ref{obs-131} and \ref{obs-304}, we can see that the jet away from the flaring region occurs earlier than the flaring event. Therefore, we first focus on the onset of the jet. 

\subsection{Simulated Evolution of the Jet }
\begin{figure}%[ht!]
\centering 
\includegraphics[width=\textwidth]{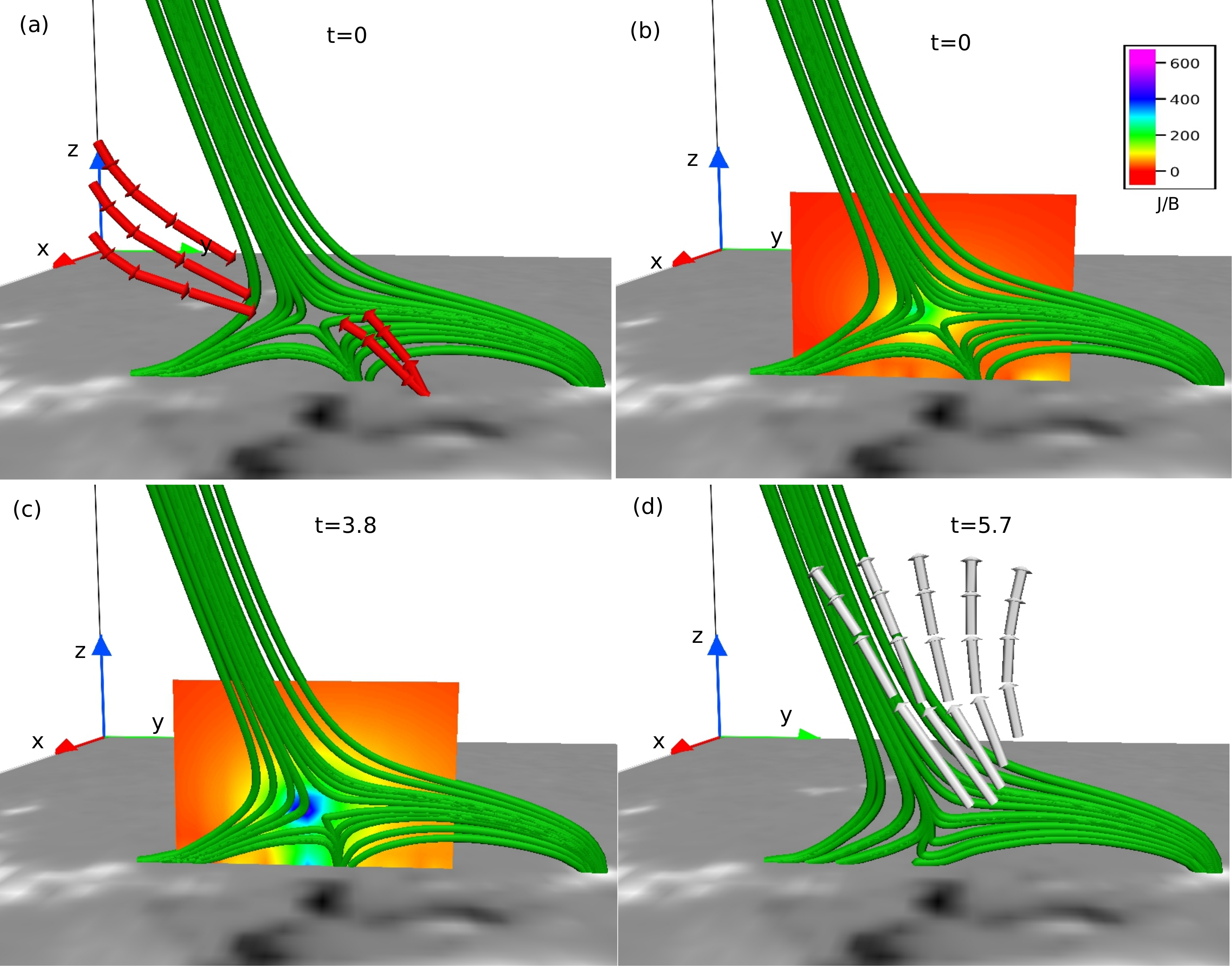}
\caption{The figure shows the evolution of the project field lines of the 3D null on a $x$-constant plane in the neighborhood of the jet. Panel (a) also shows the direction of the initial Lorentz force (red arrows). Panels (b) and (b) are overplotted with $|\mathbf{J}|/|\mathbf{B}|$. Panel (d) overlies the flow in gray-colored arrows. The bottom boundary shows Bz in grayscale.}
\label{jet-reco}
\end{figure}

To understand the onset of the jet, we investigate the reconnection at the 3D null located in the vicinity of the event (shown by the green-colored field lines in Figure \ref{extrapol}). To demonstrate the reconnection, in Figure \ref{jet-reco}, we plot the field lines of the transverse field (achieved by setting $B_x$=0 in ${\bf{B}}$), which represent the project magnetic field lines on a $x$-constant plane in the vicinity of the null. Notable is the translation of the 3D null into an X-type null in the project plane. In panel (a), the red arrows mark the direction of the initial Lorentz force near the null that is favorable to push the non-parallel field lines towards each other. As two non-parallel field lines approach in proximity, there is an enhancement in $|\mathbf{J}|/|\mathbf{B}|$, seen in panels (b) and (c), suggesting a sharp increase in the magnetic field gradient. This leads to the development of under-resolved scales which are regularized by magnetic reconnection in the simulation \citep{2016ApJ...830...80K, 2016PhPl...23d4501K}. Subsequently, the reconnection-accelerated flow is generated, as shown by gray arrows in panel (d).

\begin{figure}[ht!]
\centerline{\includegraphics[width=\textwidth]{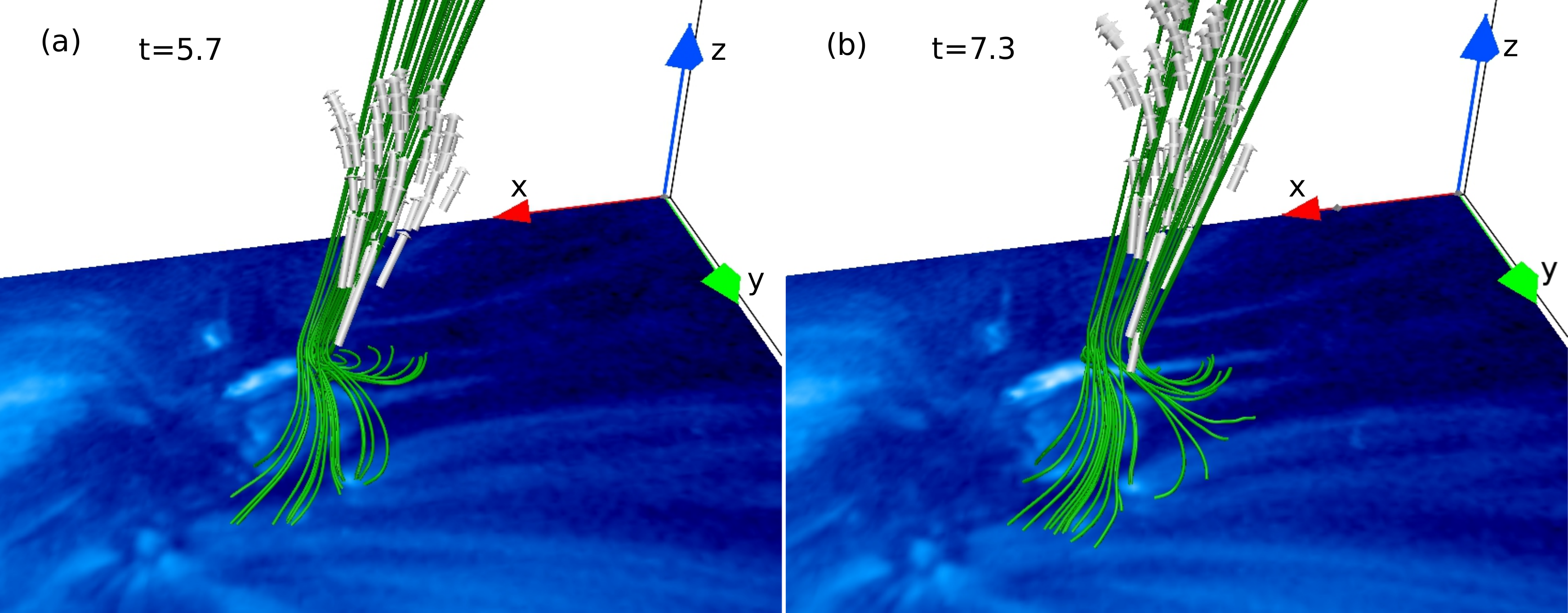}}
\caption{Panels (a) and (b) plot the post-reconnection evolution of the null (shown in Figure \ref{jet-reco}) overlaid with the jet eruption in 131 {\AA} channel. The gray arrows mark the reconnection-generated outflow which is directed along the outer spine of the null. }
\label{jet-obs}
\end{figure}

Figure \ref{jet-obs} illustrates the evolution of the 3D null along with the flow (the direction of that is represented by
the gray arrows) at t=5.7 and t=7.3. In the figure, the bottom boundary shows the contemporary AIA 131 {\AA} images which demonstrate the near match of the event with the null location. 
Notably, the flow is predominantly aligned along the outer spine of the null (see Figure \ref{jet-obs}) --- suggesting the crucial role of reconnection at the null in the onset of the event.

\subsection{Evolution of Field Lines in the Flaring Region }
\begin{figure}[ht!]
\centerline{\includegraphics[width=\textwidth]{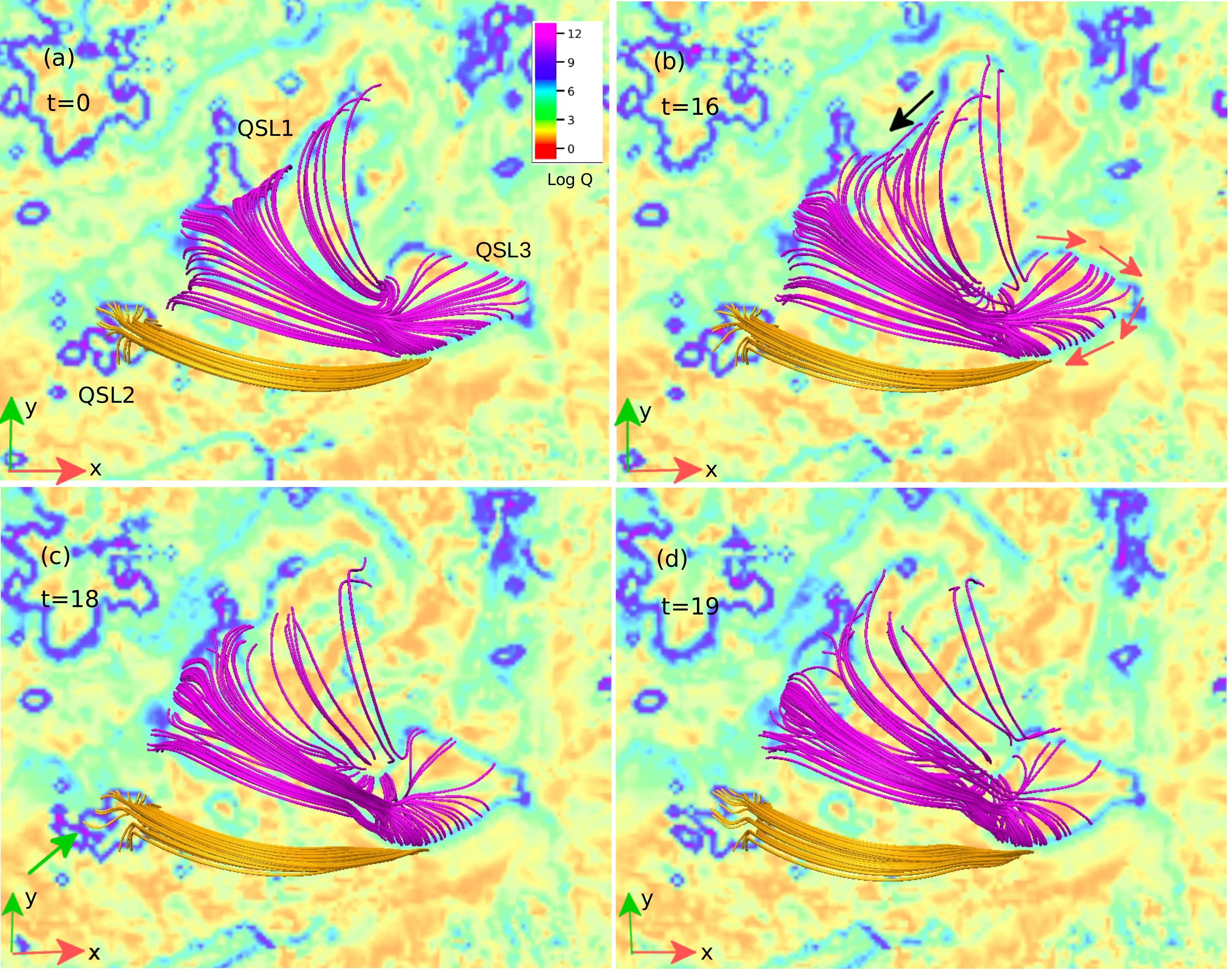}}
\caption{The figure illustrates the evolution of field lines in the flaring region overplotted with LogQ at the bottom boundary. Three distinct QSLs are marked by QSL1, QSL2, and QSL3 (panel (a)). The motion of field line footpoints is marked by a black arrow for the QSL1 (panel (b)), red arrows for the QSL3 (panel (b)), and a green arrow for the QSL2 (panel (c)).  An animation of this figure is available.}
\label{qsl-reco}
\end{figure}

Next, to explore the flare dynamics, we examine the evolution of the flaring region that is comprised of the pink and yellow-colored field lines in Figure \ref{obs-extra}. As mentioned in section \ref{obs-extra}, the pink-colored field lines represent the dome-shaped fan and the anchored spine of a 3D null. The field lines in the neighborhood of the outer spine spread from each other to form a QSL, named QSL1 (Figure \ref{qsl-reco}(a)). The fanning-out of one end of the yellow-colored loops leads to another QSL, labeled as QSL2, in the region (Figure \ref{qsl-reco}(a)). The footpoints of the dome-shaped fan also develop one more QSL, named QSL3 (Figure \ref{qsl-reco}(a)). 

To explore the possibility of reconnection at the QSLs, in Figure {\ref{qsl-reco}}, we display the time profile of the field lines overlaid with the squashing factor Q at the bottom boundary. We can notice that the motion of the field lines is such that their footpoints mostly reside on the high $Q$-values (LogQ $>$ 8) regions --- pointing toward the slipping magnetic reconnections at QSLs \citep{aulanier2006SoPh}. We have also checked the direction of plasma flow in the QSL regions, which is different from the motion of the footpoints --- further supporting the onset of slipping reconnection (not shown). Moreover, noteworthy is the motion of the footpoints in the downward direction in QSL1 as denoted by a black arrow in Figure {\ref{qsl-reco}}(b). The footpoints of the dome undergo rotational motion in QSL3 in a clockwise direction as marked by red arrows in Figure {\ref{qsl-reco}}(b). We also observe the motion of the footpoints in QSL2 (see green arrow in Figure {\ref{qsl-reco}}(c)). 

\begin{figure}[ht!]
\centerline{\includegraphics[width=\textwidth]{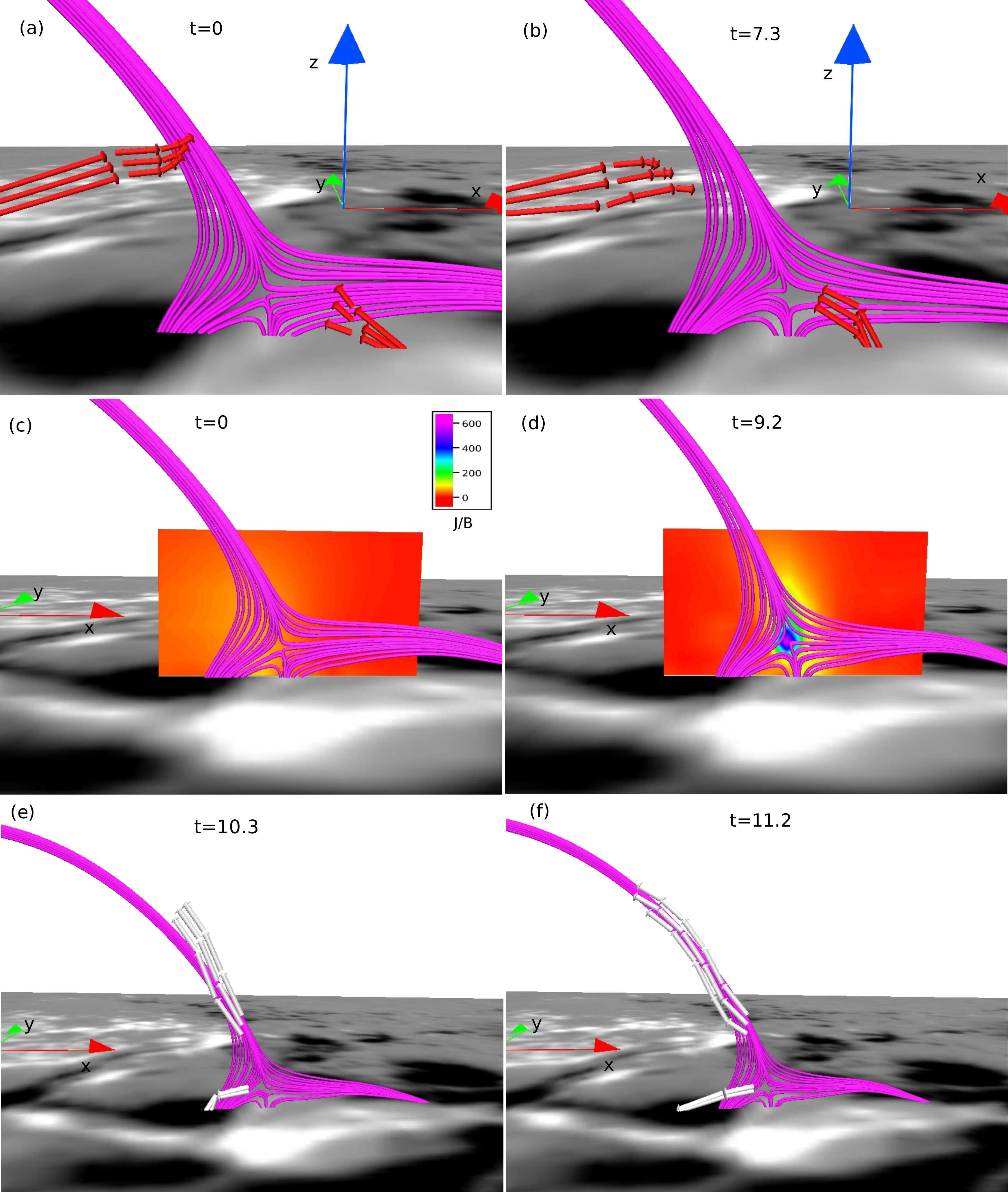}}
\caption{The figure depicts the project field lines of the 3D null (located in the flaring region) on a $y$-constant plane. Panels (a) and (b) also plot the direction of the Lorentz force (red arrows) at $t=0$ and $t=7.3$. Panels (c) and (d) are overlaid with $|\mathbf{J}|/|\mathbf{B}|$. Panel (d) overplots the flow in gray-colored arrows. The bottom boundary shows Bz. }
\label{null-reco}
\end{figure}

Furthermore, to inspect the reconnection at the 3D null in the flaring region, in Figure \ref{null-reco}, we provide 2D plots of the null by plotting the field lines of the corresponding transverse field (i.e., $B_y$=0 in ${\bf{B}}$). The field lines amount to the project magnetic field lines on a $y$-constant plane. Evident is the conversion of the 3D null into an X-type null in the plots. Panels (a) and (b) are further superimposed with the Lorentz force (the direction of which is shown by red arrows) at $t=0$ and $t=7.3$. Worthy to notice is that here the initial Lorentz force is less favorable (panel (a)) in comparison to the force in the vicinity of the jet (Figure \ref{jet-reco}(a)). This may contribute to the observed delay in the flaring event in comparison to the jet. With time, the Lorentz force becomes increasingly favorable (Figure \ref{null-reco}(b)) to bring non-parallel field lines in proximity. The resultant enhancement in the magnetic field gradient is documented by illustrating an increase in $|\mathbf{J}|/|\mathbf{B}|$ (see panels (c) and (d)).  
This ultimately leads to the onset of reconnection, which generates plasma flow. The direction of the generated flow is shown by gray-colored arrows in panels (e) and (f). The flow is found to be mostly aligned along the field lines.

\begin{figure}[ht!]
\centerline{\includegraphics[width=\textwidth]{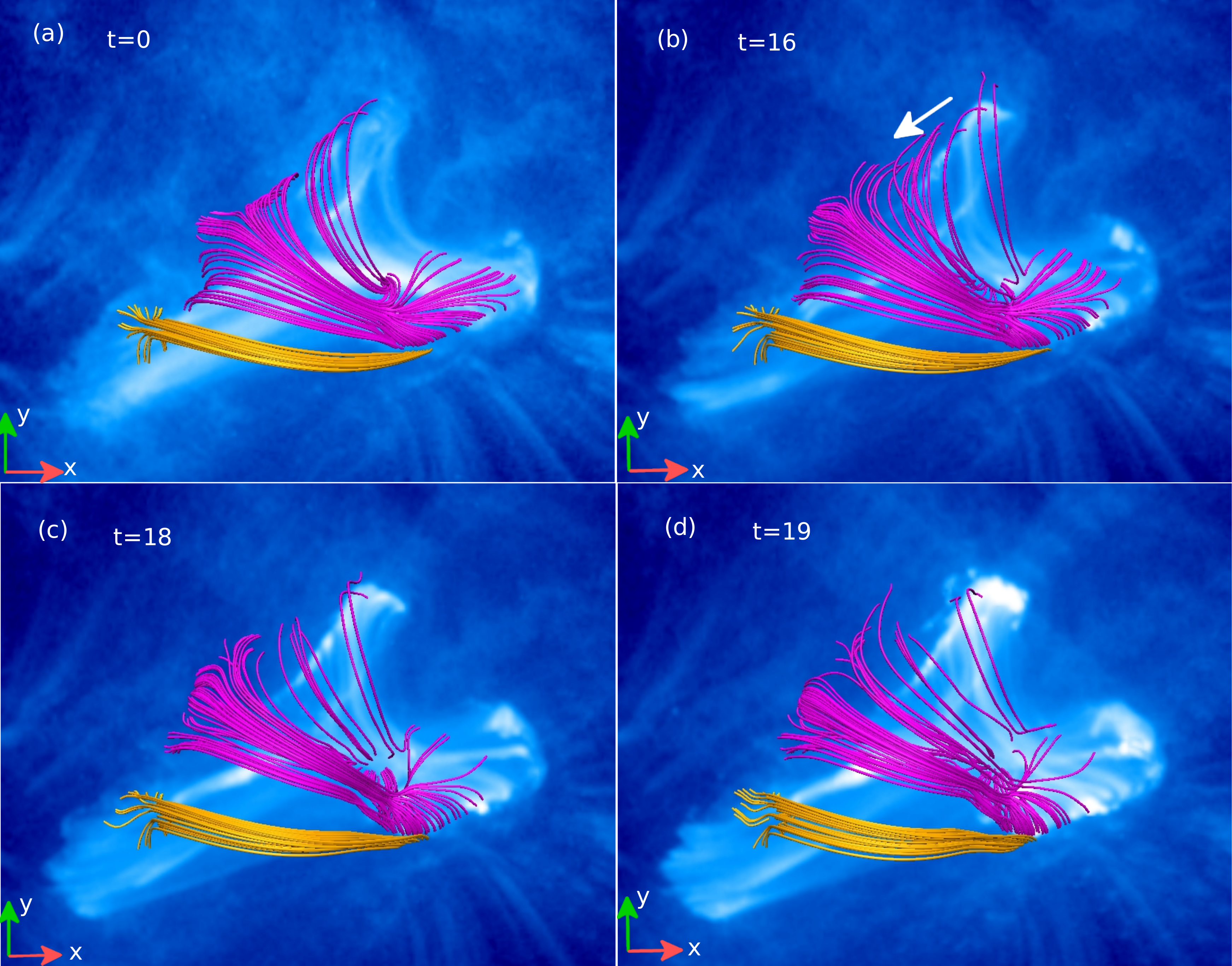}}
\caption{Superimposition of the field line evolution with co-temporal 131 {\AA} images in the flaring region. The downward motion of field lines in QSL1 and associated brightening are denoted by white arrow in panel (b). The circular-shaped brightening almost overlaps with the QSL3 (panels (c) and (d)). Also, faint brightening is observable in the QSL2. An animation of this figure is available.}
\label{131-field}
\end{figure}

\begin{figure}[ht!]
\centerline{\includegraphics[width=\textwidth]{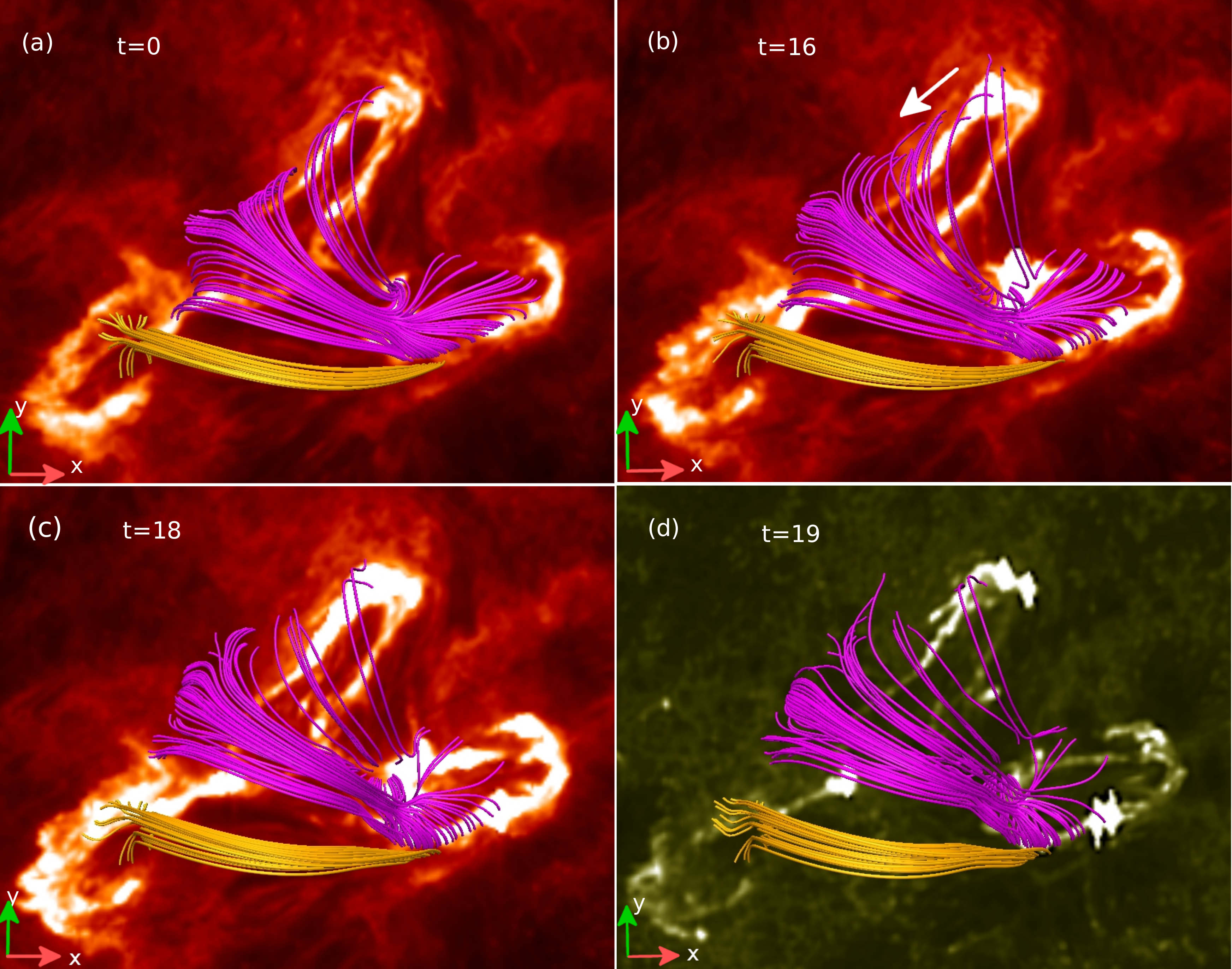}}
\caption{Similar to Figure {\ref{131-field}}, panels (a)-(c) demonstrate the field line evolution with the co-temporal 304 {\AA} images. The white arrow in panel (b) marks the similar motion of field lines in the QSL1. Panels (c) and (d) relate the field lines with the flare ribbons in 304 and 1600 channels, respectively.  An animation of panels (a)-(c) is available. }
\label{304-field}
\end{figure}

\subsection{Relation of the Field line Evolution with Observations }
To explore the correspondence between the simulated dynamics with the observations of the flaring region, in Figure \ref{131-field}, the time evolution of the field lines of the region is overlaid onto the co-temporal observations in 131 {\AA}. Importantly, the movement of the field lines in QSL1 (marked in white arrow in panel (b)) exhibits similar directionality as the downward motion of the brightening identified in observations (see panels (b)-(d)). This suggests that the underlying slipping reconnections are responsible for the brightening (BR1) in this region. Additionally, the charged particles accelerated from the 3D null on the west side can travel along the field lines and can contribute to the brightening. In the same way, the particles can also move along the field lines constituting the dome-shaped fan --- leading to the circular-shaped brightening (BR2). The slipping reconnection at QSL3 also further enhances this brightening (panels (c) and (d)). Comparatively less brightening (BR3) near the QSL2 can be attributed to the slipping reconnection at the QSL (panel (d)). The yellow-field lines of QSL2 are not found to be connected to the 3D null and, therefore, the null-point reconnection does not contribute to this brightening, which may cause less brightening.

Figure \ref{304-field} illustrates the observations of the flare in 304 {\AA} (panels (a)-(c)) and 1600 {\AA} (panel (d)) at different times. The observations are overplotted with the field lines in the flaring region. Similar to the 131 {\AA} observations, we notice an overall similarity between the downward motion of the pink field lines (denoted by a white arrow in panel (b)) in the QSL1 and the development of the elongated flare ribbon R1. This infers that the underlying slipping reconnection causes the ribbon brightening. The 3D null-point reconnection also seems to play a role in further enhancing the brightening, as the field lines are connected to the 3D null. Hence, we can imply that the slipping reconnection at the QSL1 along with the null-point reconnection are central to the formation of the ribbon R1 in 1600 {\AA} (cf. panels (d) of Figure \ref{obs-304} and Figure \ref{304-field}). At the same time, noteworthy is that the motion of the observed R1 is less farther in comparison to the footpoints of the pink field lines. This mismatch can be attributed to the limited accuracy of the initial NFFF in the weak field region of the QSL (see Figure \ref{extrapol}(a)).
  The circular ribbon R2 in 304 {\AA} (Figure \ref{304-field}(c)) and  in 1600 {\AA} (see panel (d) of Figure \ref{obs-304} and Figure \ref{304-field}) can be attributed to the null-point reconnection and the rotational motion of the field lines in QSL3 due to slipping reconnection. Finally, the semi-circular ribbon R3 in 304 {\AA} (Figure \ref{304-field}(c)) and in 1600 {\AA} (see panels (d) of Figure \ref{obs-304} and Figure \ref{304-field}) is found to appear around the QSL2. This indicates that the slipping reconnection at QSL2 may contribute to the formation of the ribbon R3. In the absence of a connection to the 3D null, null-point reconnection does not seem to contribute the ribbon R3, which also favors the less brightness in R3 in comparison to R1 and R2. It is noticeable that the spread of the foot-points of the yellow field lines is much smaller than the observed R3. This deviation also appears to be caused by the errors arose in the NFFF extrapolations of the weak field regions (Figure \ref{extrapol}(a)). In addition we would like to highlight that there is a complete absence of a twisted structure such as flux rope in the flaring region. 
 This is in agreement with the confined nature of the flaring event.

\section{Summary and Discussions}
\label{summary}
The paper aims to explain the initiation of an X-class confined flare in the NOAA AR 11155 on 2011 March 9 using multi-wavelength observations and data-constrained MHD simulation. For generating the initial magnetic field, we employ the novel non-force-free-field extrapolation technique and obtain the extrapolated field from the HMI/SDO photospheric vector magnetogram of the active region at 23:00 UT on 2011 March 9. Being non-force-free, the initial field then has Lorentz force (which becomes negligible at the coronal heights) which is crucial for spontaneously generating the dynamical evolution of the active region. During the evolution, whenever the distance between non-parallel field lines falls below the chosen grid resolution, the scales become under-resolved. The numerical model then produces locally adaptive residual dissipation and, hence, regularizes the scales by simulating magnetic reconnection.

Multi-wavelength observations of the event document three morphologically distinct brightenings; marked by the green, pink,  and yellow arrows in Figure \ref{obs-131}(d) and Figure \ref{obs-304}(c). The brightening that appears early in time (highlighted by the yellow arrows) is found to move in a downward direction with time. The additional circular and semi-circular-shaped brightenings (denoted by the green and pink arrows) start to appear. 
These brightenings are found to be associated with three distinct flare ribbons in AIA 1600 {\AA} (marked by R1, R2, and R3 in Figure \ref{obs-304}(d)). In addition to this, a jet is also observed before the initiation of the flare. The event is located away from the flaring region. 

The overall initial magnetic topology of the non-force-free extrapolated field exhibits a good match with the corresponding observations of the event, particularly in AIA 131 {\AA}. In the flaring region, the magnetic topology is characterized by a 3D null with a dome-shaped fan surface and anchored spine. Three QSLs are also identified: (a) QSL due to the spreading of the field lines in the vicinity of the spine (QSL1), (b) QSL corresponding to footpoints of the 3D null dome (QSL2), and (c) QSL associated with the fanning out of the field lines (QSL3) located in southward of the QSL1. In addition, one more 3D null is found in the vicinity of the jet. 

In the simulated dynamics, magnetic reconnection at the 3D null with an open spine initiates early in time and leads to the onset of the jet before the triggering of the flaring event. Later, in the flaring region, slipping reconnection at QSL1 and reconnection at the 3D null with anchored spine jointly appear to generate the pre-flare brightening near the QSL1 and, ultimately, lead to the development of the flare ribbon R1. The circular-shaped brightening and the associated flare ribbon R2 are proposed to be formed by the slipping reconnection at the QSL3 and the null-point reconnection. In the absence of a direct connection of field lines to the 3D null, only slipping reconnection at QSL2 contributes to the comparatively less brightening region and the corresponding flare ribbon R3.   

Overall, the MHD simulation presented in the paper successfully demonstrates the role of complex 3D magnetic structures such as null points and QSLs in the initiation of flare and further in its confinement.
Notably, the previous studies explained 
the confined flare dynamics with relatively simpler flare ribbons such as either circular \citep{Masson2009, prasad+2018apj} or parallel \citep{2016ApJ...828...62J}. In comparison, the present work advances the understanding of the flare dynamics for a complex system of circular, semi-circular, and elongated flare ribbons. Moreover, the simulation identifies two crucial factors for the confined nature of the flare:  (i) the anchored spine of the 3D null located in the flaring region, and (ii) the complete absence of the twisted structure such as flux rope. In the future, we aim to further quantify the specific roles of these factors in determining the nature of the flare. In addition, we also plan to advance the simulation by relaxing the incompressibility and including an apt physical resistivity.

%%%%%%%%%%%%%%%%%%%%%%%%%%%%%%%%%%%%%%%%%%%%%%%%%%%%%%%%%%%%%%%%%%%%%%%%%%%
\begin{acks}
We acknowledge the use of the visualization software VAPOR 
(\url{www.vapor.ucar.edu}) for generating relevant graphics. The computations were performed on the Param Vikram-1000 High Performance Computing Cluster of the Physical Research
Laboratory (PRL). Part of the computation was carried out on the computing cluster Pegasus of IUCAA, Pune, India. Data and images are courtesy of NASA/SDO and the HMI and AIA science teams. SDO/HMI is a joint effort of many teams and individuals to whom we are greatly indebted for providing the data.  We thank the referee for providing insightful suggestions and comments which led to the significant improvement of this paper.
\end{acks}

\begin{authorcontribution}
S.K., P.K., S., S.S.N. analyze the observations and simulation data. P.K. and S. prepared the plots. S.K. and S.S.N. wrote the main draft of the paper. S.A. and A.P. performed MHD simulation, calculated squashing factor, and contributed to the discussion on the results. R.B. and R.C. also contributed to the interpretation of the results. All the authors did a careful proofreading of the text and enhanced the quality of the paper.
\end{authorcontribution}

\begin{fundinginformation}
S.K., S., and R.C. would like to acknowledge the support from the DST-SERB project No. SUR\slash2022\slash000569. S.K. also acknowledges support from Patna University project No. RDC/MRP/07. 
S.S.N. acknowledges NSF-AGS-1954503, NASA-LWS-80NSSC21K0003, and
80NSSC21K1671 grants.
A.P. acknowledges the support from the Research Council of Norway through its Centres of Excellence scheme, project number 262622, as well as through the Synergy Grant number 810218 459 (ERC-2018-SyG) of the European Research Council. A.P. also acknowledges partial support from NSF award AGS-2020703.
\end{fundinginformation}

\begin{dataavailability}
The data that support the findings of this study are available from
the corresponding author upon reasonable request.
\end{dataavailability}

%\begin{declaration}
\begin{conflict}
The authors declare that they have no conflicts of interest.
\end{conflict}
%\end{declaration}

%%%%%%%%%%%%%%%%%%%%%%%%%%%%%%%%%%%%%%%%%%%%%%%%%%%%%%%%%%%%%%%%%%%%%%%%%%%
  
%%% BIBLIOGRAPHY %%%%%%%%%%%%%%%%%%%%%%%%%%%%%%%%%%%%%%%%%%%%%%%%%%%%%%%%%%%
\bibliographystyle{spr-mp-sola}
     % name your Bibtex file containing your references (.bib)
\bibliography{references}  

     % Checking: look if the file containing the ``\bibitem'' exits
     %           so check if the .bbl file exist (bibTeX compilation)
     
%\IfFileExists{\jobname.bbl}{} {\typeout{}
%\typeout{****************************************************}
%\typeout{****************************************************}
%\typeout{** Please run "bibtex \jobname" to obtain} \typeout{**
%the bibliography and then re-run LaTeX} \typeout{** twice to fix
%the references !}
%\typeout{****************************************************}
%\typeout{****************************************************}
%\typeout{}}
%%\end{article} 
\end{document}